\documentclass[twocolumn,showpacs,preprintnumbers,amsmath,amssymb,aps,prb]{revtex4}
\usepackage{graphicx}
\begin{document}

\title{
  Shapiro Steps for Skyrmion Motion on a Washboard Potential with
  Longitudinal and Transverse ac Drives} 
\author{
  C. Reichhardt
  and  C. J. Olson Reichhardt 
} 
\affiliation{
Theoretical Division,
Los Alamos National Laboratory, Los Alamos, New Mexico 87545 USA\\ 
} 

\date{\today}
\begin{abstract}
We numerically study the behavior of two-dimensional skyrmions
in the presence of a
quasi-one-dimensional
sinusoidal substrate under the influence of externally applied dc
and ac drives.
In the overdamped limit, when both dc and ac drives are aligned in the
longitudinal direction parallel to
the direction of the substrate modulation,
the velocity-force curves
exhibit classic Shapiro step features when the frequency of the ac drive
matches the washboard frequency that is dynamically generated by
the motion of the skyrmions over the substrate,
similar to previous observations in superconducting vortex systems.
In the case of skyrmions,
the additional contribution to the skyrmion motion 
from a non-dissipative Magnus force
shifts 
the location of the locking steps to higher
dc drives, and we find that the skyrmions move at an angle
with respect to the direction of the dc drive.
For a longitudinal dc drive and a perpendicular or transverse ac drive,
the overdamped system exhibits no Shapiro steps;
however, when a finite Magnus force is present we find pronounced
transverse Shapiro steps along with complex two-dimensional periodic
orbits of the skyrmions in the phase-locked regimes.
Both the longitudinal and transverse ac drives produce
locking steps whose widths oscillate
with increasing ac drive amplitude.
We examine the role of collective skyrmion interactions
and find that additional fractional locking steps
occur for both longitudinal and transverse ac drives.
At higher skyrmion densities, the system undergoes a series of
dynamical order-disorder transitions, with the skyrmions forming a
moving solid on the phase locking steps and a fluctuating dynamical
liquid in regimes between the steps.
\end{abstract}
\maketitle

\section{Introduction}
Phase locking or synchronization effects can arise in coupled oscillators 
when the different frequencies lock together over a certain range of
parameter space, an effect that
was first reported by Huygens for the synchronization of pendulum
clocks \cite{1}.
Phase locking has been extensively 
studied for numerous dynamical systems
ranging from a pair of
coupled oscillators to an entire coupled oscillator array
\cite{2,3}.
A single particle moving over a tilted one-dimensional washboard
potential can also experience phase locking when an additional ac
driving force is applied.
The substrate periodicity produces intrinsic periodic modulations
of the particle velocity in the absence of an ac drive
which increase in frequency as the magnitude of
the tilt or dc drive increases.
Addition of an external fixed-frequency ac drive produces locking regimes
in which the average dc velocity remains constant even as the magnitude
of the dc drive is increased.
The same picture can be applied to Josephson junctions,
where the analog of a velocity-force curve is
the voltage-current curve, which exhibits
a series of phase locking regions called Shapiro steps under an
applied ac drive for
single junctions \cite{4,5} 
and coupled arrays of junctions \cite{6}. 
One of the hallmarks of Shapiro steps is that the
step width oscillates as a 
function of the ac drive amplitude  \cite{4,5,6}.
Shapiro step phenomena also arise in dc and ac driven 
charge density waves \cite{7,8,9}, spin density waves \cite{10},
and Frenkel-Kontorova models 
consisting of commensurate or incommensurate arrangements of particles
moving over ordered or disordered substrates \cite{11,12,13}.
In the case of vortex motion in type-II superconductors,
Martinoli {\it et al.} reported the first observation of Shapiro steps
for dc and ac driven vortices interacting with a periodic one-dimensional (1D)
substrate created by periodic thickness modulations of the sample \cite{14,M}, 
while similar effects were observed for vortices
driven over 1D \cite{15,16} or two-dimensional (2D) \cite{17,18}
periodic substrates.
More recently, Shapiro steps 
have been found for
ac and dc driven colloidal particles 
moving over a quasi-1D periodic substrate \cite{19}.  

Shapiro steps can also occur when a lattice of collectively interacting
particles moves over a {\it random} substrate under combined dc and ac
drives.
Here, the effective elastic coupling between the particles comprising
the lattice generates an intrinsic washboard frequency that can lock to
the applied ac driving frequency.
Such steps have been studied for vortices
moving over random disorder \cite{20,21,22,23,S} or through
confined channel geometries \cite{24}.
For particles confined to 2D and moving
over a quasi-1D substrate, both the ac and dc drives must be applied
in the same direction to produce Shapiro steps;
however, for vortices moving over 2D periodic or
egg-carton substrates, it is possible
to obtain what are called transverse phase locking steps 
when the ac drive is perpendicular to the direction of the dc
drive \cite{25,26,27}.
These phase locking steps are distinct from Shapiro steps, and
their widths grow
quadratically with increasing ac amplitude
rather than showing the oscillatory behavior associated with Shapiro steps.
Phase locking effects can
also occur for overdamped particle motion in 2D periodic systems
under combinations of two perpendicular ac drives,
producing 
localized and delocalized motion as well as
rectification effects \cite{28,29,30,31,32,33}. 

In systems such as vortices and colloidal particles, an overdamped
description of the equations of motion is appropriate.  
In contrast, the skyrmions that were recently discovered in
chiral magnets have particle-like properties and many similarities
to superconducting vortices, but have the important distinction that
there is a strong non-dissipative Magnus force in their motion
\cite{34,35,36,37,38,39,40,41,42,43}. 
The skyrmions can be set into motion by an applied 
current and are observed to have a very small depinning threshold
\cite{36,37,38,39,44,45},
in part because the effectiveness of the Magnus force can be up to ten times
stronger than the dissipative force component.
The Magnus force introduces a velocity component of the
skyrmion that is perpendicular to the direction of an imposed
external force, so
a skyrmion deflects from or spirals around an attractive pinning site
rather than moving directly toward the potential minimum as
would occur for
systems governed by overdamped dynamics \cite{38,39,45,46,47,48,49}.
Since skyrmions are particle-like objects, many of their dynamical
properties can be captured using a point particle model based on a
modified Theile's equation that
takes into account repulsive skyrmion-skyrmion
interactions, the Magnus force, damping, and
substrate interactions \cite{45,A}.
Such as approach has been shown to match well 
with micromagnetic modeling \cite{45} of the
depinning of skyrmions in periodic \cite{46} and random
pinning arrays \cite{47}. 
Particle-based skyrmion models were used to describe the 
motion of skyrmions interacting with single pinning
sites \cite{48,49} as well as skyrmion motion in confined regions \cite{50}. 
Since skyrmions can easily be driven with an applied external drive they
potentially open a new class of experimentally accessible
dynamical systems where the Magnus force has a dominant effect.
It should be possible
to create various types of potential energy landscapes for skyrmions
through techniques such as
thickness modulations, periodic applied stain, controlled irradiation, 
or spatially periodic doping.
An open question is 
how known phase locking phenomena would be affected by the inclusion
of a Magnus force,
and whether new types of phase locking effects
might appear that are absent in overdamped systems.
Skyrmions also have potential for various spintronic applications
\cite{51} which would require the skyrmions to move in a controlled manner,
so an understanding of skyrmion phase locking dynamics
could be useful for producing new methods for precision control
of skyrmion motion.

\begin{figure}
\includegraphics[width=3.5in]{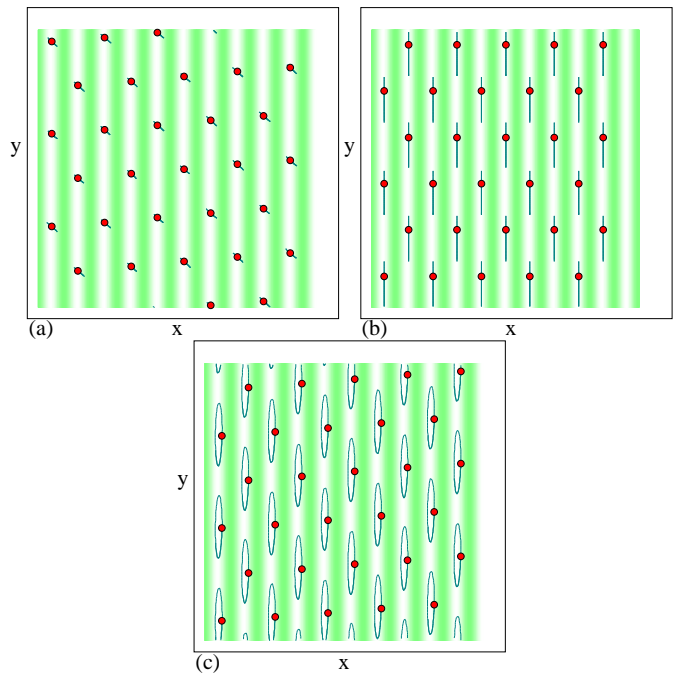}
\caption{
(Color online) Skyrmions (red dots) at a density of
$\rho_s=0.001$ on a
periodic quasi-1D substrate with $A_p=1.0$.
The darker regions are potential maxima
and the lighter regions are potential minima, while
lines indicate the skyrmion trajectories. 
(a) For an ac drive $F^{ac}_x=1.0$ applied in the longitudinal
or $x$-direction at $\alpha_{m}/\alpha_{d} = 1.0$
and $F^{dc}=0$,
the skyrmions oscillate in 1D paths at a $45^{\circ}$
angle to the $x$ axis.
(b) For an ac drive $F^{ac}_y=0.75$ applied in the transverse or
$y$-direction with $F^{dc}=0$ at $\alpha_{m}/\alpha_{d} = 0.0$
or the overdamped limit, the skyrmions
move in 1D paths along the $y$ direction.
(c) The same as in (b) with
$\alpha_{m}/\alpha_{d} = 1.0$, where the skyrmions
form elliptical 2D counterclockwise orbits.    
}
\label{fig:1}
\end{figure}

In this work we examine Shapiro steps for skyrmions moving over a
quasi-1D periodic washboard substrate
similar to the geometry considered by Martinoli {\it et al.}
for vortices moving over substrates with a periodic
thickness modulation
\cite{14,M,15}.
We consider the motion of single
skyrmions and collectively interacting skyrmions 
over a periodic substrate, as illustrated in
Fig.~\ref{fig:1}.
Here the longitudinal or $x$-direction is aligned with the
direction of the periodicity of the substrate,
while the $y$-direction corresponds to the transverse
direction.
In this geometry in the overdamped limit, Shapiro steps occur when
the dc ($F^{dc}$) and ac ($F^{ac}$) driving forces
are both applied along the $x$-direction.
When a finite Magnus force is present,
we find
that phase locking continues
to occur; however, the
net skyrmion motion is rotated at an angle with respect to the
substrate periodicity direction.
As the magnitude of the Magus force prefactor
is increased, the width and number of steps for a fixed
dc drive gradually decreases, and the steps shift to higher values of $F^{dc}$. 
In the overdamped limit, if the ac drive is applied
in the transverse or $y$-direction, no   
Shapiro steps occur, but for a finite Magnus force
a new set of Shapiro steps
can arise, with the width and the number of resolvable steps in the
velocity-force curve increasing with increasing 
Magnus force prefactor.
On a phase-locked step,
the skyrmion motion takes the form of intricate
periodic 2D orbits.
As a function of the ac amplitude, we show that the
widths of the
phase locking steps for both longitudinal and transverse ac driving
exhibit the oscillatory
behavior associated with Shapiro steps.
  
We note that the Shapiro steps we observe
are different than the previously studied 
phase locking motion for
purely dc-driven skyrmions moving over a
2D periodic substrate \cite{46}.
In the latter case, the phase locking was associated with
a directional locking effect
in which the skyrmion motion locks to certain symmetry
directions of the substrate potential. 
In the present study there is no phase locking without an ac drive.  

\section{Simulation and System}   

We consider a 2D system of size $L \times L$ with periodic
boundary conditions in the $x$ and $y$ directions
containing $N_s$ skyrmions
at a density of $\rho_s=N_s/L^2$.  Single ($N_s=1$) or multiple
skyrmions
interact with a quasi-1D periodic sinusoidal potential with a periodicity
direction running along the $x$ direction,
as illustrated
in Fig.~\ref{fig:1}.
The equation of motion for a single skyrmion $i$ 
with velocity ${\bf v}_i=d{\bf r}_i/dt$ moving in the $x-y$ plane
is
\begin{equation}
  \alpha_{d}{\bf v}_{i} + \alpha_{m}{\hat z}\times {\bf v}_{i} = {\bf F}^{ss}_{i} +  {\bf F}^{sp}_{i} 
  + {\bf F}^{dc} + {\bf F}^{ac} .
\end{equation}
Here ${\bf r}_{i}$ is the location of the skyrmion and
$\alpha_{d}$ is the prefactor of the damping force
that aligns the skyrmion velocity in the direction
of the net external forces.
The second term is the Magnus force with prefactor $\alpha_{m}$,
which rotates the velocity into the direction {\it perpendicular} to the
net external forces.
In order to maintain a constant magnitude of the skyrmion velocity 
we impose the constraint $\alpha_{d}^2 + \alpha^2_{m} = 1$
and vary the relative importance of the Magnus force to the
damping force by changing the ratio $\alpha_{m}/\alpha_{d}$.
In the overdamped limit $\alpha_{m} = 0.0$, while for skyrmions
$\alpha_{m}/\alpha_{d}$ can be ten or larger \cite{39,45}.

The skyrmion-skyrmion interaction force is
${\bf F}^{ss}_{i} = \sum_{j=1}^{N_{s}}K_{1}(R_{ij}){\bf \hat r}_{ij}$
where $R_{ij} = |{\bf r}_{i} - {\bf r}_{j}|$, 
${\hat {\bf r}_{ij}} = ({\bf r}_i - {\bf r}_{j})/R_{ij}$, and
$K_{1}$ is the modified Bessel function. This 
interaction is repulsive and falls off exponentially for large $R_{ij}$.
For most of this work we remain
in the limit where skyrmion-skyrmion interactions are
weak so that we can consider the dynamics of
a single skyrmion; however, we show that most of our
results are robust under the inclusion of
skyrmion-skyrmion interactions.
The substrate force
${\bf F}^{sp}_i  = -\nabla U(x_i) {\bf \hat x}$ arises from a washboard potential
\begin{equation}
U(x)  = U_{0}(\cos(2\pi x_i/a)) 
\end{equation}
where $x_i={\bf r}_i \cdot {\bf \hat x}$,
$a$ is the periodicity of the substrate, and
we define the substrate strength to be $A_{p} = 2\pi U_{0}/a$.
Unless otherwise noted, we take $A_p=1.0$.
The dc driving term ${\bf F}^{dc}=F^{dc}{\bf \hat x}$
is slowly increased in magnitude to avoid any transient effects. 
The ac driving
term is
either ${\bf F}^{ac}_{x}= F^{ac}_{x}\cos(\omega t){\bf \hat x}$
for transverse driving or
${\bf F}^{ac}_{y} = F^{ac}_{y}\cos(\omega t){\bf \hat y}$
for perpendicular driving. 

We measure the time-averaged skyrmion velocities
in the $x$ direction
$\langle V_x \rangle =
\sum_{i=1}^{N_s}2\pi\langle {\bf v}_i \cdot {\bf \hat x} \rangle/N_s\omega a$
and $y$-direction 
$\langle V_y\rangle =
\sum_{i=1}^{N_s}2\pi\langle {\bf v}_i \cdot {\bf \hat y} \rangle/N_s\omega a$. 
Here,
due to the periodicity of the 
substrate, phase locked steps occur when the
skyrmions travel integer multiples of 
the substrate periodicity $na$ during each ac drive cycle,
allowing us to label the steps
$n = 0$ for the pinned phase and $n = 1, 2...$ for the 
higher order steps.  
We focus on the two ac frequencies
$\omega = 8\times 10^{-4}$ inverse simulation time steps
for the longitudinal ac driving
and $\omega = 1.6\times 10^{-3}$ inverse simulation time steps
for the transverse ac driving,
and use a substrate lattice constant of $a = 3.272$.
 
We use two different driving protocols as illustrated in Fig.~\ref{fig:1}.
For longitudinal driving, we have
\begin{equation}
{\bf F}^{\rm drive} = F^{dc}{\bf \hat x} + F^{ac}_x\cos(\omega t){\bf \hat x},
\end{equation} 
corresponding to the conditions under which
Shapiro steps arise for an overdamped system.
For transverse driving, we have
\begin{equation}
{\bf F}^{\rm drive} = F^{dc}{\bf \hat x} + F^{ac}_y\cos(\omega t){\bf \hat y}, 
\end{equation}
which would produce no Shapiro steps in the overdamped limit.

In Fig.~\ref{fig:1}(a) we show the skyrmion
trajectories 
for $F^{ac}_{x} = 1.0$, $F^{dc} = 0.0$,
$\alpha_{m}/\alpha_{d} = 1.0$, and a skyrmion    
density of $\rho_s=0.001$.
In this case the skyrmions are pinned and form a
triangular lattice that is commensurate with the substrate. 
The ac drive causes the skyrmions to oscillate in the potential
minima; however, their motion is not strictly in 
the $x$-direction but is tilted 
at an angle of $\theta = 45^{\circ}$ with respect to the $x$ direction
due to the Magnus force,
which induces a velocity component perpendicular to the
ac driving direction.
In the absence
of a substrate, a dc or ac drive applied in the
$x$-direction causes the skyrmions to move at an angle
$\theta = \arctan(\alpha_{m}/\alpha_{d})$
with respect to the driving direction,
so that in the overdamped limit of $\alpha_{m} = 0.0$ the skyrmion 
moves parallel to the direction of the net external driving force.  
In Fig.~\ref{fig:1}(b), we rotate the direction of the ac drive to
be in the transverse direction with $F^{ac}_y=0.75$ and $F^{dc}=0$ for
a sample with
$\alpha_{m}/\alpha_{d} = 0.0$. 
In this case the skyrmion motion follows strictly 1D paths
aligned with the $y$-direction that pass through 
the potential minima of the substrate.
For $\alpha_{m}/\alpha_{d} = 1.0$, as shown in Fig.~\ref{fig:1}(c),
the skyrmions rotate in
counterclockwise
elliptical patterns,
showing that the Magnus force can induce
$x$-direction motion even when the drive is applied only in the
$y$-direction.
In the absence 
of the substrate the ac drive would produce only
1D trajectories at an angle with respect to the $y$-axis.
This highlights the fact that 
the Magnus force affects how the skyrmions move when
interacting with forces induced by the substrate.    

\section{Longitudinal ac Driving}

\begin{figure}
\includegraphics[width=3.5in]{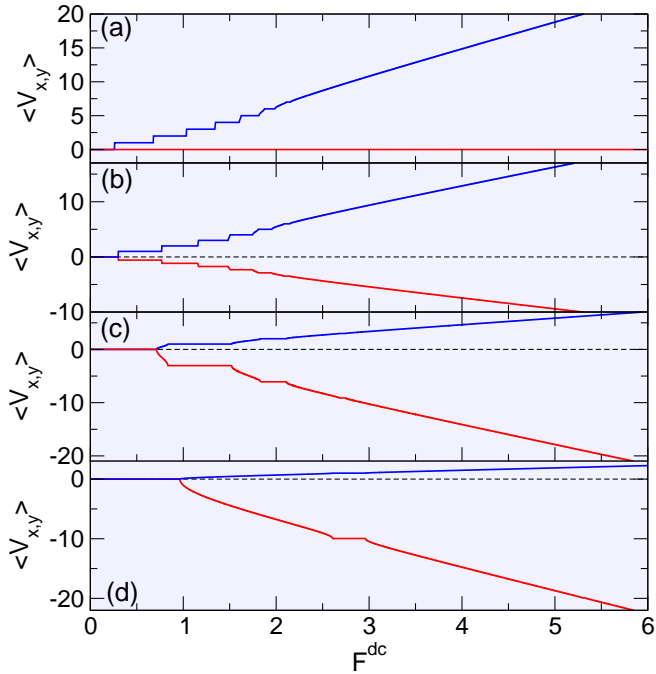}
\caption{ $\langle V_{x}\rangle$ (upper blue curves) and
$\langle V_{y}\rangle$ (lower red curves)
vs $F^{dc}$ for the system in Fig.~\ref{fig:1}(a) in the
single skyrmion limit at $F^{ac}_{x} = 1.0$.
(a) In the overdamped limit of $\alpha_{m}/\alpha_{d} = 0$,
$\langle V_y\rangle = 0$ 
and a series of steps appear in $\langle V_x\rangle$
indicating phase locking. 
(b) At $\alpha_{m}/\alpha_{d} = 0.58$,
$\langle V_{y}\rangle$ is finite. 
(c) $\alpha_{m}/\alpha_{d} = 3.042$ and (d) $\alpha_{m}/\alpha_{d} = 9.962$
show the increase of
skyrmion motion in the
direction transverse to the substrate and the shift in the locking phases.}    
\label{fig:2}
\end{figure}

We first consider the case
illustrated in Fig.~\ref{fig:1}(a) of ac
driving in the longitudinal direction.  We conduct a series
of simulations for increasing
$\alpha_{m}/\alpha_{d}$ and focus on the single skyrmion limit.
In 
general we find that the Shapiro steps we observe remain robust when finite
skyrmion-skyrmion interactions are included; however, additional features
can arise for varied fillings when the skyrmion structure is
incommensurate with the substrate, as we discuss in Section V.
In Fig.~\ref{fig:2}(a) we plot $\langle V_{x}\rangle$
and $\langle V_{y}\rangle$ versus $F^{dc}$ for
the system in Fig.~\ref{fig:1}(a) at $F^{ac}_{x} = 1.0$
in the overdamped
limit of $\alpha_{m}/\alpha_{d} = 0$.
Here $\langle V_{y}\rangle = 0$ while 
$\langle V_{x}\rangle$ shows a series of steps indicative of the phase
locking.
These features are similar to those observed for other overdamped
systems moving over quasi-1D periodic substrates such as
vortices in type-II superconductors moving over 
quasi-1D substrate modulations. 
In Fig.~\ref{fig:2}(b), when
$\alpha_{m}/\alpha_{d} = 0.58$, both
$\langle V_{y}\rangle$ and $\langle V_{x}\rangle$ are finite and have
a ratio of $|\langle V_{y}\rangle/\langle V_{x}\rangle| \approx 0.58$.
Here the phase locking is still occurring, but the
intervals of $F^{dc}$ in which the phase locking steps appear
are shifted.
Figure~\ref{fig:2}(c) shows that at
$\alpha_{m}/\alpha_{d} = 3.042$,
both $|\langle V_{y}\rangle|$ 
and some of the step widths have increased
in size, and there are no clear regions
between the steps where no phase locking is occurring. 
In Fig.~\ref{fig:2}(d), at
$\alpha_{m}/\alpha_{d} = 9.962$, there is only a single phase locking step.

\begin{figure}
\includegraphics[width=3.5in]{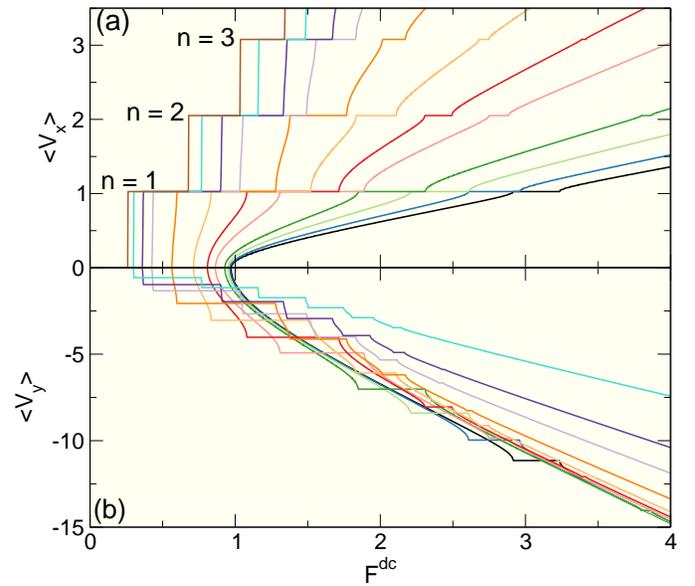}
\caption{ 
(a) $\langle V_x\rangle$ vs $F^{dc}$ at $A_{p} = 1.0$ 
for 
$\alpha_{m}/\alpha_{d} = 0.0$ (brown),
0.577 (light blue),
0.98 (dark purple),
1.33 (light purple),
2.06 (dark orange),
3.042 (light orange),
4.0 (dark red),
4.92 (light red),
7.0 (dark green),
8.407 (light green),
9.962 (dark blue),
and $11.147$ (black),  
from left to right.
Here $\langle V_{x}\rangle$ exhibits quantized values 
corresponding to specific steps.
(b) The corresponding values of
$\langle V_{y}\rangle$ vs $F^{dc}$, which contains steps
that are not quantized.
}
\label{fig:3}
\end{figure}

\begin{figure}
\includegraphics[width=3.5in]{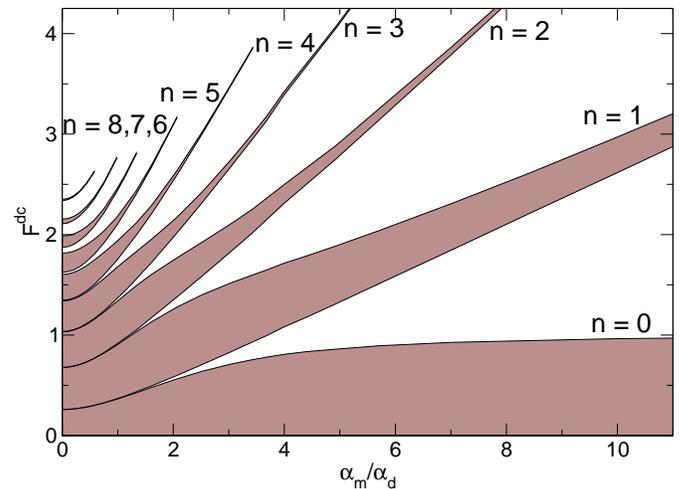}
\caption{The regions of phase locking for the $n = 0$ to $n = 8$ steps
as a function of $F^{dc}$ and $\alpha_{m}/\alpha_{d}$.
The width of the steps is reduced and the steps shift to
higher values of $F^{dc}$ with
increasing $\alpha_{m}/\alpha_{d}$. 
}
\label{fig:4}
\end{figure}
      
To more clearly demonstrate the behavior
of the steps for varied $\alpha_{m}/\alpha_{d}$,
in Fig.~\ref{fig:3}(a) we plot $\langle V_{x}\rangle$ versus
$F^{dc}$ for $\alpha_{m}/\alpha_{d}$ ranging
from $0.0$ to $11.147$, with the evolution of the first three
locking steps $n=1$ to 3 highlighted.
For a given value of $n$, the step in $\langle V_x\rangle$ has a fixed
value regardless of the choice of $\alpha_m/\alpha_d$, and each step
shifts to higher values of $F^{dc}$ with increasing
$\alpha_{m}/\alpha_{d}$.
The corresponding 
$\langle V_{y}\rangle$ versus $F^{dc}$ plot in Fig.~\ref{fig:3}(b)
shows that the steps in $\langle V_y\rangle$
are not quantized in integer multiples of $2\pi/a\omega$.
The quantization of the $\langle V_{x}\rangle$ arises from the
periodicity of the substrate in the $x$-direction, and since the
$y$-direction has no periodicity, there is no quantization of
$\langle V_y\rangle$.
In Fig.~\ref{fig:4} we highlight the evolution of the
widths of the $n = 0$ through $n=8$ steps as a function of
$F^{dc}$ and $\alpha_{m}/\alpha_{d}$ at $F^{ac}_{x} = 1.0$.
At $\alpha_{m}/\alpha_{d} = 0$, the largest number of phase
locking steps can be resolved.
We observe two general trends as
$\alpha_{m}/\alpha_{d}$ increases.
First, for $n > 3$, the widths of the locking regions decrease 
and the intervals  of $F^{dc}$ over which the locking occurs
shift to higher values of $F^{dc}$, with the magnitude of
this shift increasing
with increasing $n$.
Second, the width of the $n = 1$, 2, and $3$ steps initially
increases for
increasing $\alpha_{m}/\alpha_{d}$ 
before reaching a maximum and then decreasing
again.
The width of the $n = 0$ step reaches a maximum
with increasing $\alpha_m/\alpha_d$ and then saturates. 
The shift in the locations of the phase locking regions arises because
the angle at which the skyrmions move with respect to the $x$-axis increases
with increasing $\alpha_{m}/\alpha_{d}$, causing the skyrmions to spend larger
intervals of time interacting with the repulsive portion of the
substrate potential.  As a result, higher values of $F^{dc}$ must be
applied to  
cause the skyrmion to translate in the $x$-direction at
larger $\alpha_m/\alpha_d$.

\begin{figure}
\includegraphics[width=3.5in]{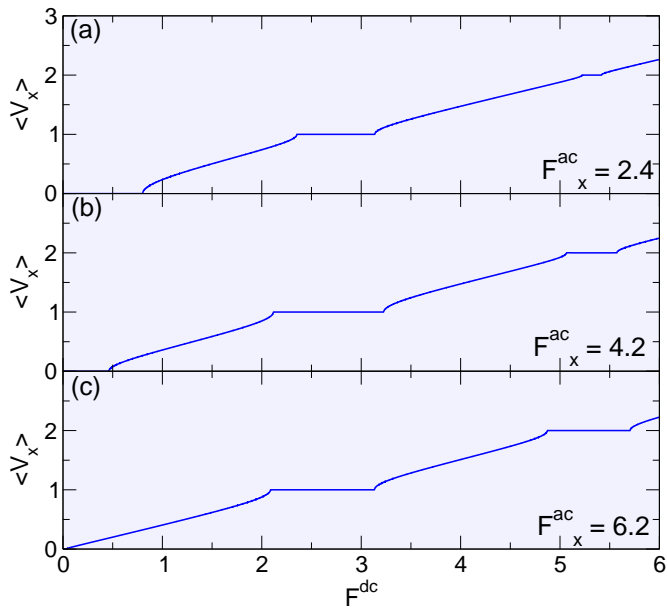}
\caption{ $\langle V_{x}\rangle$ vs
$F^{dc}$ for $\alpha_{m}/\alpha_{d} = 9.962$.
(a) $F^{ac}_{x} = 2.4$.
(b) $F^{ac}_x=4.2$.
(c) $F^{ac}_x=6.2$.
}
\label{fig:5}
\end{figure}

\begin{figure}
\includegraphics[width=3.5in]{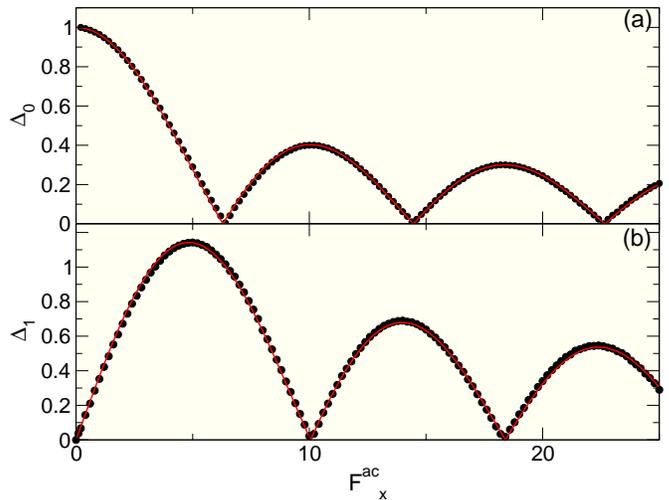}
\caption{(a) The width $\Delta_0$ of the $ n = 0$ step vs $F^{ac}_x$
for the system in
Fig.~\ref{fig:5} at
$\alpha_{m}/\alpha_{d} = 9.962$.
The solid line is a fit to the $|J_{0}|$ Bessel function.
(b) The width $\Delta_{1}$ of the $n=1$ step vs $F^{ac}_{x}$ for
the same system.
The solid line is a fit to the $|J_{1}|$ Bessel function.
In each case the width of step $n$ shows an oscillation
of the form of the Bessel function
$|J_{n}(F^{ac}_{x})|$, which is characteristic of Shapiro step
phase locking.   
}
\label{fig:6}
\end{figure}

We next determine if the phase locking steps at
high Magnus force prefactor are of the Shapiro type.
In Fig.~\ref{fig:5} we plot
$\langle V_{x}\rangle$ versus
$F^{dc}$ for $\alpha_{m}/\alpha_{d} = 9.962$ at
$F^{ac}_{x} = 2.4$, 4.2, and $6.2$
to show the variation in the widths of the $n = 0$, $n=1$, and $n=2$ steps. 
In Fig.~\ref{fig:6} we plot the widths $\Delta_0$ and $\Delta_1$ of the
$n=0$ and $n=1$ steps, respectively, versus 
$F^{ac}_{x}$.
Each step shows the characteristic oscillation expected for Shapiro steps,
where 
the width of step $n$ is proportional to
$|J_{n}(F^{ac}_x)|$,
where $J_n$ is the $n$th-order Bessel function \cite{5}.
The solid lines in Fig.~\ref{fig:6}(a,b)
are fits to $|J_{0}|$ and $|J_{1}|$, respectively.
The higher order steps obey similar fits.
This indicates that in the Magnus-dominated limit, Shapiro step
phase locking is occurring. 

\section{Transverse ac Driving}

\begin{figure}
\includegraphics[width=3.5in]{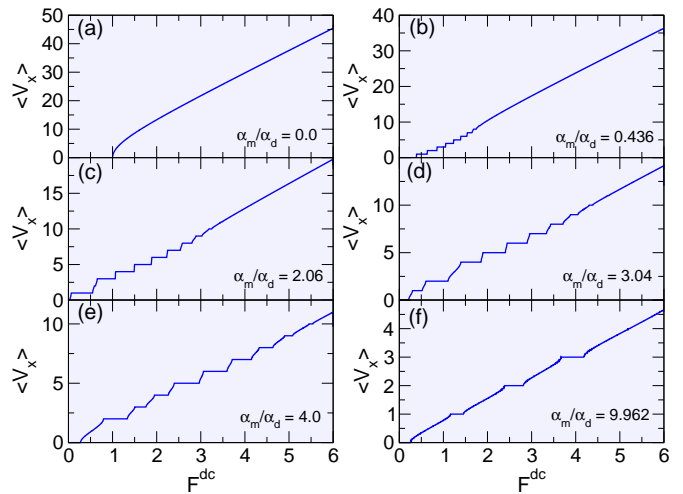}
\caption{ $\langle V_{x}\rangle$ vs $F^{dc}$ for a system
with dc driving in the $x$-direction and ac driving $F^{ac}_{y} = 1.0$
in the $y$-direction.
(a) At $\alpha_{m}/\alpha_{d} = 0.0$, there are no steps
in $\langle V_x\rangle$.
(b) At $\alpha_{m}/\alpha_{d} = 0.436$, steps are present.
(c) $\alpha_{m}/\alpha_{d} = 2.06$.
(d) $\alpha_{m}/\alpha_{d} = 3.04$.
(e) $\alpha_{m}/\alpha_{d} = 4.0$.
(f) $\alpha_{m}/\alpha_{d} = 9.962$.    
}
\label{fig:7}
\end{figure}

We next consider the case illustrated in Fig.~\ref{fig:1}(b,c),
where the ac drive is applied transverse to the direction of the
substrate potential.
In the overdamped limit of $\alpha_{m}/\alpha_{d} = 0$,
such a drive causes the skyrmion to
oscillate in the $y$-direction as shown in Fig.~\ref{fig:1}(b),
and when a finite dc drive is applied in the longitudinal direction,
a single washboard oscillation
frequency in the $x$-direction is generated by the motion of the skyrmion
over the periodic substrate.  Since only one frequency is
present, there is no coupling between two frequencies,
so mode locking does not occur.
When the Magnus force is finite,
the transverse ac drive
induces an oscillating velocity component in the
longitudinal or $x$-direction
as well as in the $y$-direction, as illustrated
in Fig.~\ref{fig:1}(c), so that it is possible for the dc-induced washboard
frequency to couple to 
the transverse ac frequency and generate
a transverse Shapiro step.     
In Fig.~\ref{fig:7} we plot $\langle V_{x}\rangle$ vs $F^{dc}$
for a single skyrmion moving with $F^{ac}_{y} = 1.0$.
At $\alpha_m/\alpha_d=0$, shown 
in Fig.~\ref{fig:7}(a), there are no steps in $\langle V_{x}\rangle$,
indicating the lack
of phase locking, while the corresponding $\langle V_{y}\rangle = 0.$
Depinning occurs at the threshold value $F_{c}$ of $F_c = A_{p} = 1.0$.
Figure~\ref{fig:7}(b) shows that at $\alpha_{m}/\alpha_{d} = 0.436$, 
the depinning threshold has dropped substantially to $F_{c} = 0.4$
and a series of steps 
are now visible for $0.4 < F^{dc} < 2.0$,
indicating that phase locking is occurring.
For $\alpha_{m}/\alpha_{d} > 0.0$,
$\langle V_y\rangle$ is finite and
the $|\langle V_{y}\rangle|$ versus $F^{dc}$ curve has
exactly the same form as $\langle V_{x}\rangle$ versus $F^{dc}$,
but the magnitude of $|\langle V_y\rangle|$ is
multiplied by $\alpha_{m}/\alpha_{d}$.
In Fig.~\ref{fig:7}(c,d) we plot
$\langle V_{x}\rangle$ versus $F^{dc}$ for samples with
$\alpha_{m}/\alpha_{d} = 2.06$ and $3.04$, respectively.
Here, 
the widths of the locking steps increase with increasing
$\alpha_m/\alpha_d$ and the step locations are shifted to
higher values of $F^{dc}$. 
In samples with $\alpha_{m}/\alpha_{d} = 4.0$ and $9.962$,
as shown in Fig.~\ref{fig:7}(e,d), respectively,
the steps extend 
out to larger values of $F^{dc}$, and the non-phase locking regions
between the steps are also extended. 
The steps in $\langle V_x \rangle$ once again occur
at quantized values of $n a\omega/2\pi$ due to the 
periodicity in the $x$-direction, while the steps
in $\langle V_y\rangle$ do not have quantized values.

\begin{figure}
\includegraphics[width=3.5in]{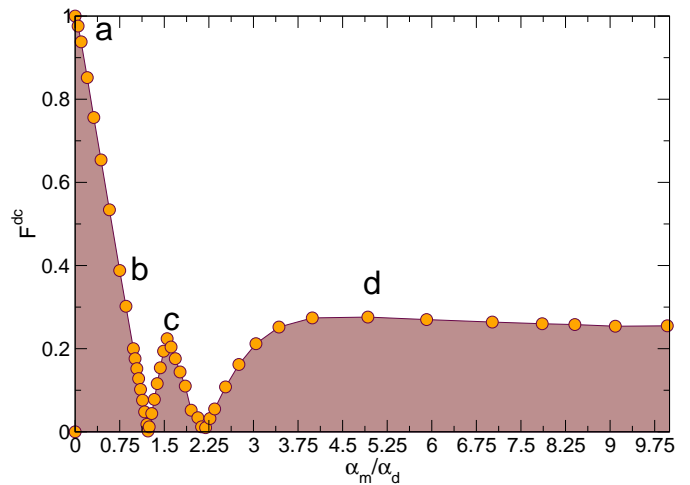}
\caption{ 
The location of the upper edge of the $n = 0$ step
as a function of $F^{dc}$ and $\alpha_{m}/\alpha_{d}$
for the system in Fig.~\ref{fig:6} with a transverse ac
drive of $F^{ac}_{y} = 1.0$.
Here there are several local minima and maxima that are associated with
changes in the skyrmion orbits, as
shown in Fig.~\ref{fig:9} at the points marked a-d.   
}
\label{fig:8}
\end{figure}

In Fig.~\ref{fig:8} we plot the location of the upper edge of
the $n=0$ step as a function of $F^{dc}$ and $\alpha_m/\alpha_d$ for the
system shown in Fig.~\ref{fig:6} with $A_p=1.0$.  This is equivalent to the
threshold depinning force $F_c$.
Here
$F^{dc}/A_{p}=1.0$ at $\alpha_{m}/\alpha_{d} = 0.0$ and it
decreases to zero at $\alpha_{m}/\alpha_{d} = 1.226$.
There is a local maximum in $F^{dc}/A_p$ at $\alpha_{m}/\alpha_{d} = 1.55$,
followed by another minimum near $\alpha_{m}/\alpha_{d} = 2.2$ and
a broad plateau for higher values of 
$\alpha_{m}/\alpha_{d}$.
This oscillatory behavior in the $n = 0$ step width is absent for
longitudinal ac driving, 
as shown in Fig.~\ref{fig:5} where $F_{c}$ exhibits only monotonic behavior.
The dips and 
maxima in $F_{c}$ for the transverse ac driving are
associated with transitions in the shape of the skyrmion orbits 
during a single ac drive cycle for increasing $\alpha_{m}/\alpha_{d}$.

\begin{figure}
\includegraphics[width=3.5in]{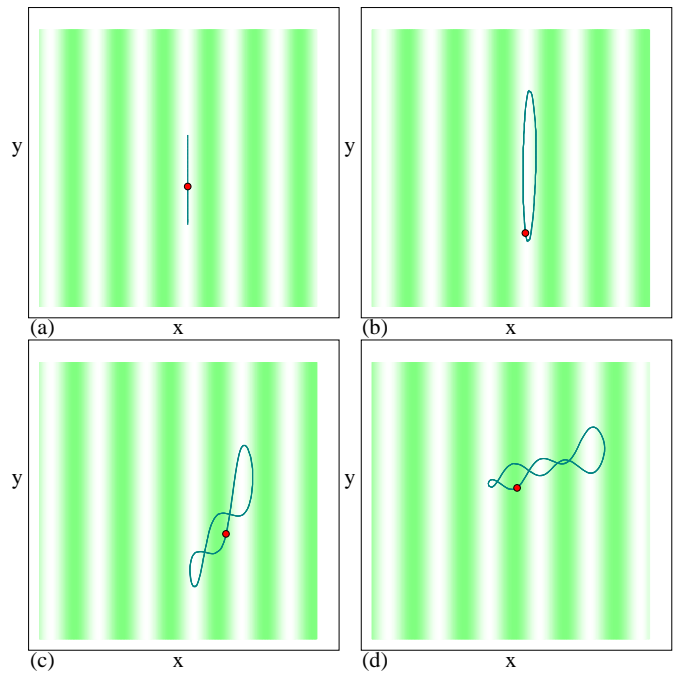}
\caption{
Skyrmions (red dots), potential maxima (darker regions), potential
minima (lighter regions), and skyrmion trajectories (lines)
in a portion of the system in Fig.~\ref{fig:8} along the $n = 0$ step
at the points labeled (a-d) in Fig.~\ref{fig:8}.
(a) At $\alpha_{m}/\alpha_{d} = 0.0$ for $F^{dc} = 0.1$,
there is 1D motion in the $y$-direction. 
(b) $\alpha_{m}/\alpha_{d} = 0.75$ at $F^{dc} = 0.2$.
(c) At $\alpha_{m}/\alpha_{d} = 1.54$ and $F^{dc} = 0.1$, the skyrmion
moves between two potential minima.
(d) At $\alpha_{m}/\alpha_{d} = 4.92$ and $F^{dc}=0.2$,
the skyrmion moves between three
potential minima.  
}
\label{fig:9}
\end{figure}

In Fig.~\ref{fig:9} we illustrate the skyrmion trajectories
in a subsection of the system
on the $n = 0$ step at the points labeled (a-d) in Fig.~\ref{fig:8}.
Figure~\ref{fig:9}(a) shows that for $\alpha_{m}/\alpha_{d} = 0$
and $F^{dc} = 0.1$, the skyrmion moves in a 1D path in
the $y$-direction along the potential minimum.
At
$\alpha_{m}/\alpha_{d} = 0.75$ and $F^{dc} = 0.2$,
in Fig.~\ref{fig:9}(b),
the skyrmion forms an elliptical orbit that is confined within a
single potential trough. 
On the local maximum in the $n=0$ step marked point c in Fig.~\ref{fig:8},
at $\alpha_{m}/\alpha_{d} = 1.54$ and $F^{dc} = 0.1$,
the skyrmion forms a more complicated 2D orbit that has
three lobes.  In a single ac drive cycle the skyrmion translates
back and forth by two substrate
lattice constants. 
The dip in $F_{c}$ at $\alpha_{m}/\alpha_{d} = 1.226$ shown in
Fig.~\ref{fig:8} corresponds to the point at which the 
skyrmion orbit transitions from being confined in one potential
minimum to traversing two potential minima. 
Above the second local minimum  
at $\alpha_{m}/\alpha_{d} = 2.2$ in Fig.~\ref{fig:8}, the skyrmion
orbit becomes even more complex, as illustrated in
Fig.~\ref{fig:9}(d) for 
$\alpha_{m}/\alpha_{d} = 4.92$ and $F^{dc} = 0.2$.
The skyrmion
now moves between three substrate potential minima
in a single ac drive cycle.
The local minimum in the $n = 0$ step width at $\alpha_{m}/\alpha_{d} = 2.2$ 
then corresponds 
to the transition in the skyrmion motion from traversing two substrate
minima to traversing three substrate minima.
For higher values of $\alpha_{m}/\alpha_{d}$, additional minima in $F_{c}$
could occur that would be correlated with orbits traversing four or more
substrate minima. 
We expect that additional substrate minima would be resolvable
in samples with a smaller substrate lattice constant $a$.

\begin{figure}
\includegraphics[width=3.5in]{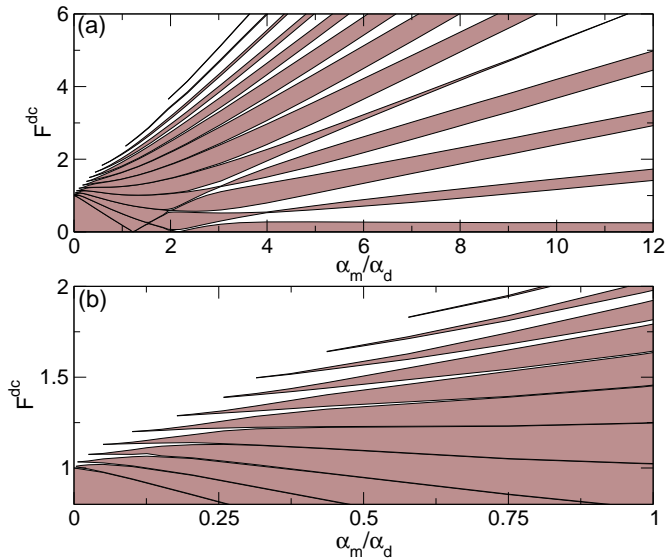}
\caption{(a) Evolution of the regions in which the
$n = 0$, 1, 2, 3, 4, 5, 6, 7, 8, 9, 10, and $11$ steps (from bottom
to top) appear as a function of $F^{dc}$ and 
$\alpha_{m}/\alpha_{d}$ for $F^{ac}_{y} = 1.0$.
Increasing the Magnus force produces enhanced phase locking.
(b) A blowup of panel (a) in the region of small $\alpha_{m}/\alpha_{d}$
showing that the 
steps vanish as $\alpha_{m}/\alpha_{d}$ goes to zero.  
}
\label{fig:10}
\end{figure}

\begin{figure}
\includegraphics[width=3.5in]{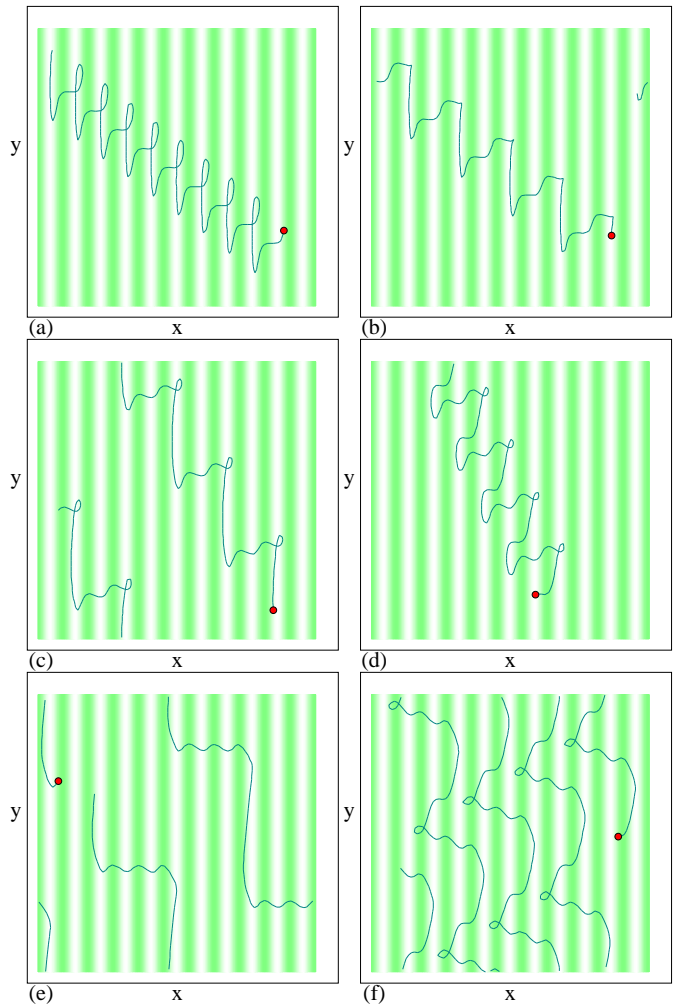}
\caption{Skyrmions (red dots), potential maxima (darker regions),
potential minima (lighter regions), and skyrmion trajectories (lines)
for the system in Fig.~\ref{fig:9}.
(a) $n = 1$ orbit at $\alpha_{m}/\alpha_{d} = 0.75$ and $F^{dc} = 0.6$.
(b) $n = 2$ orbit at $\alpha_{m}/\alpha_{d} = 0.75$ and $F^{dc} = 0.7$.
(c) $n = 2$ orbit at $\alpha_{m}/\alpha_{d} = 1.55$ and $F^{dc} = 0.5$.
(d) $n = 1$ orbit at $\alpha_{m}/\alpha_{d} = 2.06$ and $F^{dc} = 0.3$.  Here
the skyrmion translates by 2$a$ in the positive $x$ direction
followed by $a$ in the negative $x$ direction for
a net transport by a distance $a$ in the $x$-direction 
during each ac cycle.
(e) $n = 3$ orbit at $\alpha_{m}/\alpha_{d} = 2.06$ and $F^{dc} = 0.67$.
(f) $n = 1$ orbit at $\alpha_{m}/\alpha_{d} = 5.92$ and
$F^{dc} = 0.67$.   
}
\label{fig:11}
\end{figure}

In Fig.~\ref{fig:10}(a) we highlight the regions of phase locking as a
function of $F^{dc}$ and $\alpha_{m}/\alpha_{d}$ for steps $n=0$ through
$n=11$ for the system in Fig.~\ref{fig:7}.
When $\alpha_{m}/\alpha_{d} = 0$,
all the steps with $n \geq 1$ vanish, as illustrated in
Fig.~\ref{fig:10}(b) where we plot the regime
$0 \leq \alpha_{m}/\alpha_{d} \leq 1.0$.
As the Magnus force increases, a larger number
of steps can be resolved.
In general, the step widths increase with
increasing $\alpha_{m}/\alpha_{d}$;
however, certain steps such as $n = 1$, 2, and $3$ show
step width oscillations.
In the case of longitudinal ac driving, the skyrmion orbits along the
different locking steps are always 1D in nature.  In contrast, the orbits
are much more complicated for transverse ac driving. 
In Fig.~\ref{fig:11}(a) we show the $n = 1$ skyrmion orbit from
Fig.~\ref{fig:10} at 
$\alpha_{m}/\alpha_{d} = 0.75$ and $F^{dc} = 0.6$.
The skyrmion translates in the positive $x$-direction and negative
$y$-direction, making an angle close to
$\theta = \arctan(\alpha_{m}/\alpha_{d}) = 36.9^{\circ}$
with the $x$-axis.
During a single 
orbit the skyrmion passes through a loop and translates by one
lattice constant in the $x$-direction. 
Figure~\ref{fig:11}(b) illustrates the $n=2$ orbit for
$\alpha_{m}/\alpha_{d} = 0.75$ at $F^{dc} = 0.7$, where the skyrmion
translates two lattice constants 
in the $x$-direction per ac cycle.
On the $n=2$ step
for $\alpha_{m}/\alpha_{d} = 1.55$, and $F^{dc} = 0.5$,
shown in Fig.~\ref{fig:11}(c),
the skyrmion 
moves at a steeper angle of $\theta = 57.1^{\circ}$
from the $x$-axis. 
In Fig.~\ref{fig:11}(d), which shows
the $n = 1$ orbit at $\alpha_{m}/\alpha_{d} = 2.06$
and $F^{dc} = 0.3$, during a single ac drive cycle the skyrmion initially moves
$2a$ in the positive $x$-direction before moving $a$ in the negative
$x$-direction, producing a net translation in the $x$-direction
of a distance $a$ per ac cycle. 
Figure~\ref{fig:11}(e) shows
the $\alpha_{m}/\alpha_{d} = 2.06$ system in the $n = 3$ orbit
at $F^{dc} = 67$, where the skyrmion translates by $3a$ in a single ac cycle. 
On the $n=1$ step
at $\alpha_{m}/\alpha_{d} = 4.92$ and $F^{dc} = 0.67$,
plotted in Fig.~\ref{fig:11}(f),
the skyrmion
moves $3a$ in the positive $x$-direction
during the first portion of the ac drive cycle
followed by $2a$ in the negative $x$-direction during the second portion
of the ac drive cycle,
producing a net translation of $a$
in the $x$ direction during a single ac cycle.
We observe similar orbits for the other values of $n$, and find
that the net angle of the skyrmion motion with respect to
the $x$ axis increases with increasing $\alpha_{m}/\alpha_{d}$.  

\subsection{Dependence on Substrate Strength and ac Amplitude}

\begin{figure}
\includegraphics[width=3.5in]{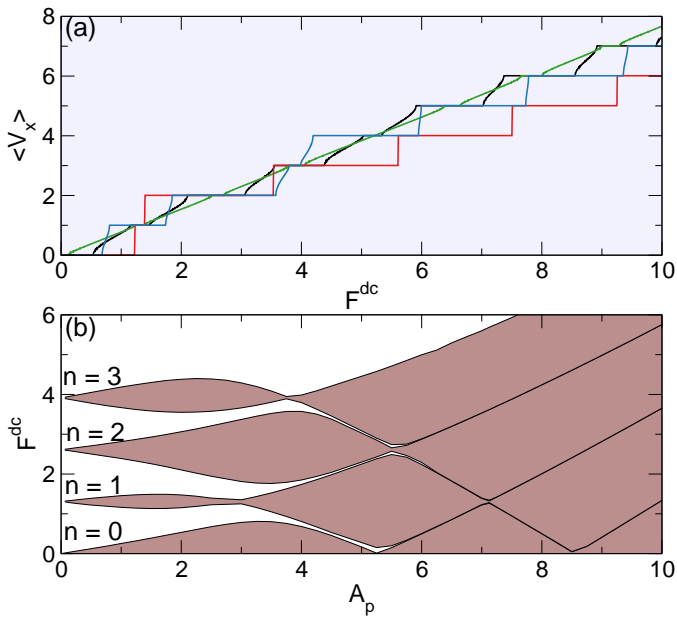}
\caption{ 
(a) $\langle V_{x}\rangle$ vs $F^{dc}$ for $F^{ac}_{y} = 1.0$,
$\alpha_{m}/\alpha_{d} = 9.962$, and $A_{p} = 0.5$ (black),
$2.0$ (green), 4.0 (blue),  and $7.0$ (red). 
(b) The evolution of the
$n=0$, 1, 2, and 3 step widths as a function of $F^{dc}$ and
$A_{p}$
for the system in panel (a). 
}
\label{fig:12}
\end{figure}

We next consider the effect of the substrate strength on the
transverse locking steps at 
$\alpha_{m}/\alpha_{d} = 9.962$ and $F^{ac}_{y} = 1.0$.
In Fig.~\ref{fig:12}(a) we plot 
$\langle V_{x} \rangle$ versus $F^{dc}$
for $A_{p} = 0.5$, 2.0, 4.0, and $7.0$.
At the lower values of $A_{p}$, the phase locking 
steps decrease in width, and the steps completely vanish when $A_{p}= 0$. 
This is highlighted in
Fig.~\ref{fig:12}(b) where we plot the
widths of the $n = 0$, 1, 2, and $3$ steps as a
function of $F^{dc}$ and $A_{p}$. 
The width of the locking
regions oscillates with increasing $A_{p}$,
and for $A_{p} > 8.0$ all the locking phases 
shift linearly to higher values of $F^{dc}$  
with increasing $A_{p}$. 
The step width oscillations arise due to variations in
the number of potential minima through which the skyrmion
orbit passes during a single ac drive cycle,
similar to what was observed for fixed $A_{p}$ and
varied $\alpha_{m}/\alpha_{d}$. 
This result shows that the transverse phase locking is
a generic feature that appears in both the strong and 
weak substrate regimes, and that it is more pronounced for stronger
substrates.

\begin{figure}
\includegraphics[width=3.5in]{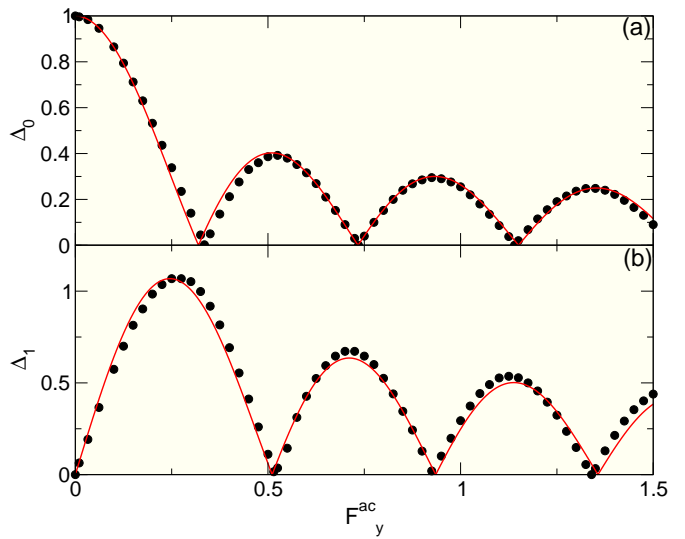}
\caption{(a) The width $\Delta_0$ of the $ n = 0$ step vs
$F^{ac}_{y}$ for the system in Fig.~\ref{fig:12} at
$\alpha_m/\alpha_d=9.962$ and $A_p=1.0$.  The solid line is a
fit to the $|J_0|$ Bessel function.  
(b) The width $\Delta_{1}$ of the $n=1$ step vs $F^{ac}_{y}$ for
the same system.  The solid line is a fit to the $|J_{1}|$ Bessel function.  
}
\label{fig:13}
\end{figure}

We next examine the dependence of the step widths
at a fixed $A_{p}$ on the ac driving amplitudes,
as shown in Fig.~\ref{fig:13} where we plot $\Delta_{0}$ and
$\Delta_{1}$ versus $F^{ac}_{y}$ for
$A_{p} = 1.0$ and $\alpha_{m}/\alpha_{d} = 9.962$.
The solid lines are fits to 
$\Delta_{n} \propto |J_{n}(F^{ac}_{y})|$,
indicating that the transverse phase locking steps are also
of the Shapiro step type,
similar to the longitudinal phase locking steps.    

\section{Collective Effects}

\begin{figure}
\includegraphics[width=3.5in]{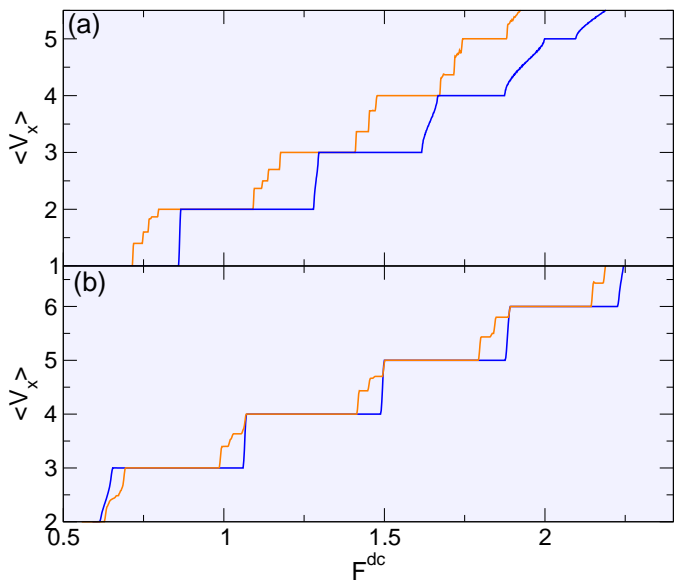}
\caption{$\langle V_{x}\rangle$ vs $F^{dc}$
at $\alpha_{m}/\alpha_{d} = 2.0$.
(a) ac driving in the $x$-direction with $F^{ac}_{x} = 1.0$
for a single skyrmion (dark blue line) and a sample containing
multiple skyrmions at a density of $\rho_s=0.04$ (light orange line),
showing that fractional
phase locking steps can arise.
(b) The same for ac driving in the $y$-direction at $F^{ac}_{y}=1.0$.   
}
\label{fig:14}
\end{figure}

We next consider assemblies of interacting skyrmions for the
system shown in Fig.~\ref{fig:1}.  In general, when
the skyrmion density is commensurate with the substrate and
the skyrmions can form a triangular lattice, skyrmion-skyrmion
interactions cancel and we find
the same types of phase locking observed in the
single skyrmion systems.
For incommensurate fillings
where dislocations are present or when the skyrmion
structure becomes distorted or anisotropic in the pinned phase,
we find that it is possible
for additional fractional phase locking to occur
between the integer phase locking steps.  These
fractional locking steps
occur when a portion of the skyrmions are locked to step $n$
and the remainder of the skyrmions are locked to step $n-1$.
In Fig.~\ref{fig:14}(a) we
plot $\langle V_{x}\rangle$ versus
$F^{dc}$ for a system with ac driving in the $x$-direction
at $\alpha_{m}/\alpha_{d} = 2.06$ and
$F^{ac}_{x} = 1.0$ to compare the results for a single skyrmion
with a system at a skyrmion density of $\rho_s=0.04$.
There are no fractional steps in the single skyrmion system;
however, when interacting skyrmions are present we find
fractional steps $n/m$, where $n$ and $m$ are integers. 
Figure~\ref{fig:14}(b) shows the same system for ac driving in the
$y$-direction, where the same types of fractional steps arise.    
The fractional steps appear at incommensurate fields
when it is possible to have two effective particle species in the
system.
One species is commensurate and the other is associated with 
interstitials, dislocations, or vacancies.
In overdamped systems such as superconducting vortices
moving over 2D periodic substrates,
similar integer steps for individual or non-interacting vortices
appear at commensurate matching fillings while
additional fractional locking steps arise    
at non-matching fields \cite{18}. 

At much higher skyrmion densities and for sufficiently strong substrate
strengths, the pinned skyrmion structures 
become highly anisotropic due to the confinement 
in the 1D pinning rows.
In the moving phase just above depinning, the effectiveness of the pinning
is partially reduced
and the repulsive skyrmion-skyrmion
interactions favor a more uniform structure.
The competition between skyrmion-skyrmion and skyrmion-substrate
interactions produces  a series of order-disorder transitions in the moving 
state.  On
the phase-locked steps, the skyrmions form an ordered moving anisotropic
lattice and
travel in a synchronized fashion,
while between the
phase locking steps the skyrmions adopt a more isotropic or liquid like
configuration.

\begin{figure}
\includegraphics[width=3.5in]{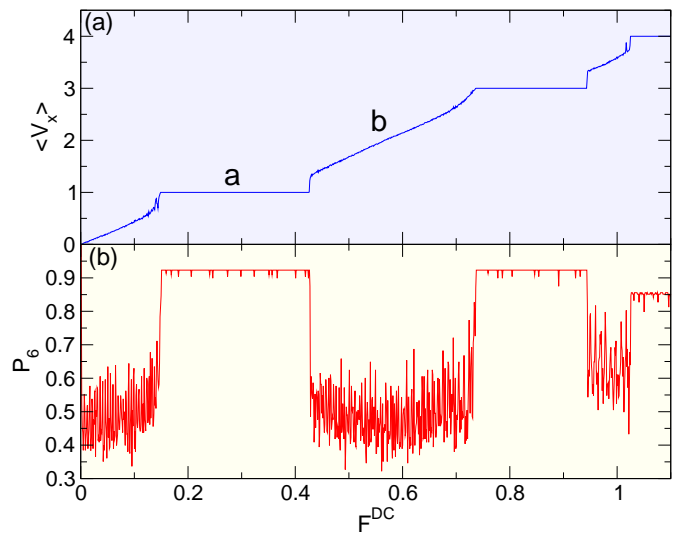}
\caption{(a) $\langle V_{x}\rangle$ vs $F^{dc}$ at
$F^{ac}_{y} = 1.0$ and $\alpha_{m}/\alpha_{d} = 2.06$
for a sample with
a skyrmion density of $\rho_s=0.4$.
(b) The fraction of six-fold coordinated particles $P_{6}$
vs $F^{dc}$ for the same system showing that 
along the phase-locked steps the skyrmions form a much more ordered state.  
}
\label{fig:15}
\end{figure}

\begin{figure}
\includegraphics[width=3.5in]{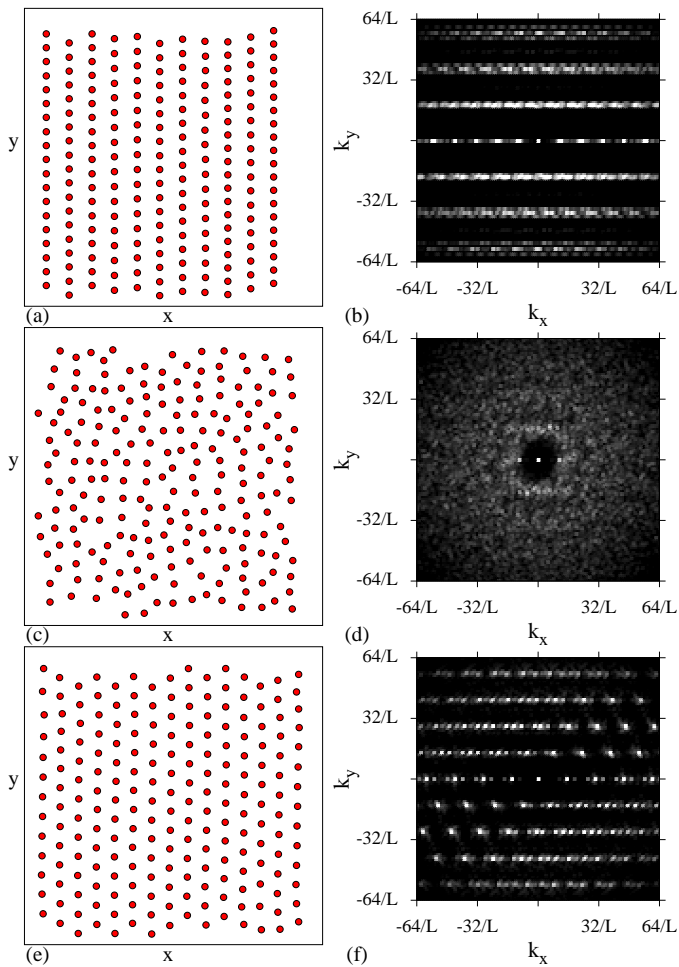}
\caption{ (a,c,e) The real space positions of the skyrmions from
Fig.~\ref{fig:15}
and (b,d,f)  the corresponding structure factors $S(k)$.
(a,b) The $n = 1$ phase locked step at $F^{dc} = 0.25$
from the point labeled {\bf a} in Fig.~\ref{fig:15}(a)
shows a partially ordered anisotropic structure. 
(c,d) On the non-step region at $F^{dc}=0.6$ labeled {\bf b} in
Fig.~\ref{fig:15}(a), the skyrmions form a disordered liquid like structure. 
(d,e) On a non-step region at $F^{dc} = 9.0$,
the system forms a moving lattice.   
}
\label{fig:16}
\end{figure}

In Fig.~\ref{fig:15}(a) we plot 
$\langle V_{x}\rangle$ versus $F^{dc}$ for a sample with
$F^{ac}_{y} = 1.0$, $\alpha_{m}/\alpha_{d} = 2.06$,
and a skyrmion density of $\rho_s=0.4$, showing the $n=1$, 3, and 4 phase
locking steps.
Figure~\ref{fig:15}(b) illustrates the corresponding
fraction of six-fold coordinated skyrmions
$P_6=N_s^{-1}\sum_{i=1}^{N_s}\delta(6-z_i)$, where $z_i$ is the coordination
number of skyrmion $i$ obtained from a Voronoi construction.
On the phase locking steps, $P_6$ increases to $P_6=0.92$,
while between the steps $P_{6} \approx 0.5$ on average 
and shows strong fluctuations.
In Fig.~\ref{fig:16}(a,b) 
we show the real space locations of the skyrmions
and the corresponding structure factor $S(k)$ 
on the $n = 1$ step at $F^{dc} = 0.25$ from Fig.~\ref{fig:15}.
The skyrmions are all moving together and form
a partially ordered but anisotropic lattice.
Even though the system is anisotropic,
most of the skyrmions have six neighbors,
so that $P_{6} \approx 0.9$.
Figure~\ref{fig:16}(c,d) shows the same sample at
$F^{dc} = 0.6$, corresponding to the non-phase locking region 
labeled {\bf b} in Fig.~\ref{fig:15}.
Here the skyrmions form a disordered structure that is less
anisotropic than the phase locked state.
We observe similar sets of dynamical order-disorder transitions
between step and non-step regions
for increasing $F^{dc}$ and find
similar effects for ac driving in the $x$-direction.
Studies in overdamped systems of collections of interacting vortices 
also show that the vortices are more ordered
and exhibit suppressed noise fluctuations
in a phase locked region \cite{22,S}. 
At higher $F^{dc}$, the effectiveness of the substrate gradually
diminishes, the phase locking steps disappear, and
the skyrmions can reorder into a more uniform moving crystal state as
shown in Fig.~\ref{fig:16}(e,f) at $F^{dc} = 9.0$.
Similar dynamical 
reordering to a triangular lattice for high drives
has been observed for skyrmions interacting
with random pinning \cite{47} as well
as for vortices driven over random pinning arrays \cite{52,53}.        
These results show that Shapiro steps for skyrmions
interacting with a periodic substrate are a robust feature that occurs
for a variety of skyrmion densities and substrate strengths.
The change in the skyrmion lattice structure as the system passes
in and out of phase locked states as a driving current is swept
could be observed using neutron scattering or noise measurements.

\section{Summary}

We have analyzed Shapiro steps for skyrmions interacting
with periodic quasi-one-dimensional substrates in the presence of 
combined dc and ac drives, with a specific focus on the role of the Magnus
force in the dynamics.
When the dc and ac drives are both applied in the longitudinal direction,
which is aligned with the substrate periodicity,
phase locking occurs, and as the role of the Magnus force increases, the
phase locking steps gradually reduce in width and shift to higher values
of the driving force.
The skyrmions move at an angle to the direction of the external
dc drive that increases as the contribution of the Magnus force increases.
When the ac drive is applied perpendicular to the dc drive and the substrate
periodicity direction, there is no phase locking in the overdamped limit;
however, when there
is a finite Magnus force, phase locking can occur.  On the  phase locked steps
the skyrmions move in intricate 
two-dimensional periodic orbits.
We map out the evolution of the phase locked regions for the
transverse and longitudinal ac driving 
for varied contribution of the Magnus force, ac driving amplitudes,
and substrate strength.
When collective interactions between skyrmions are introduced,
fractional Shapiro steps can appear.
For strong substrate strengths and higher skyrmion densities,
both longitudinal 
and transverse phase locking steps occur that are
associated with dynamically induced transitions between an
ordered anisotropic solid on the steps
to a fluctuating liquid state in the non-phase locked regimes. Such transitions
could be observed with neutron scattering.

\acknowledgments
This work was carried out under the auspices of the 
NNSA of the 
U.S. DoE
at 
LANL
under Contract No.
DE-AC52-06NA25396.

\end{document}